# Factorized Boolean representations for efficient quantum synthesis

Mehul Shah[1]*, Robert Fiszer[1], and Marek Perkowski[1]

[1]Department of Electrical and Computer Engineering, Portland State University, Portland, OR, USA

*Corresponding author: mehul@pdx.edu. R.F.: fizzmaister@gmail.com. M.P.: h8mp@pdx.edu

**Quantum algorithms promise advantages beyond classical reach, but running them on error-corrected hardware requires translating Boolean specifications into reversible circuits, and the resources that translation demands determine what is executable. Established methods minimize a Boolean expression and map it to a circuit, assuming the minimized form is best. Here we show that minimized expressions retain algebraic structure minimization cannot reach, arising from containment and complementary-polarity relationships among their terms, and that extracting it yields circuits cheaper to execute despite having more operations. The decisive quantity is not a circuit's operation count but the control count of its widest operation, a superlinear cost; extracting shared factors trades a few wide operations for many narrow ones and reduces qubit count. Across benchmarks and oracles from quantum search and factoring algorithms, at the representation level the transformation never increases either cost measure, a guarantee from its construction. Translation to an executable circuit returns part of that advantage, since auxiliary lines must be uncomputed, yet the factorized circuit still left a leading circuit-level optimizer reaching lower final counts, and faster, than unaided. The representation of a computation is therefore itself a resource, optimizable before compilation and distinct from both logic minimization and circuit-level optimization.**

Quantum computing offers computational paradigms for problems intractable on classical architectures, with algorithms such as Grover search and Shor factorization demonstrating advantages arising from interference and superposition[1,2]. Realizing those advantages on fault-tolerant hardware, however, requires transforming Boolean descriptions into reversible circuits, and the scalability of the resulting machines depends not only on algorithmic development but on the complexity of the circuit representations underneath.

The dominant cost in fault-tolerant computation is that of non-Clifford operations, particularly T gates. Each T gate consumes a distilled magic state, and magic-state distillation occupies a large fraction of both the physical qubits and the execution time of projected surface-code machines. The number of T gates a synthesis produces therefore determines whether a computation is executable at all, not merely how efficiently it runs. The scale involved is substantial: state-of-the-art estimates for factoring a 2,048-bit RSA modulus require on the order of 2.6 billion Toffoli gates[3], so reducing the resource cost of the arithmetic primitives from which such circuits are assembled is a problem of direct practical consequence.

Synthesis of Boolean transformations into reversible circuits has consequently become a central component of quantum compilation. Compact Boolean representations of this kind also underpin applications beyond circuit synthesis, including machine learning and knowledge discovery[4], oracle construction for search and constraint satisfaction[5,6], Hamiltonian formulations of Boolean optimization[7,8], and quantum state machine synthesis[9,10].

Existing approaches have investigated reversible logic synthesis, Toffoli network optimization, and quantum cost reduction through improved decomposition[11-14]. These methods optimize the reversible circuit after a Boolean representation has already been chosen, leaving an underexplored question: whether the logical representation itself contains structure that can be exploited before quantum realization.

Boolean minimization has traditionally focused on reducing expression size, through sum-of-products, exclusive-sum-of-products and Reed-Muller forms, by eliminating redundant terms and literals[15-19], with efficient heuristics developed for both binary and multiple-valued inputs[20-22]. Minimizing the number of terms or literals, however, does not minimize implementation complexity. A compact representation may still conceal hierarchical relationships among its product terms, and those relationships determine what the reversible mapping can achieve. Existing minimization therefore captures only part of the available computational structure.

## Gate arity governs implementation cost

A Toffoli gate with n controls cannot be executed directly and must be decomposed into elementary operations, the number of which grows superlinearly with n. Under the Maslov cost model an n-control gate is charged $2^{n+1} - 3$ elementary operations, so a five-control gate costs an order of magnitude more than a two-control gate, and total circuit cost is dominated by the few widest gates rather than by how many gates are present.

Conventional synthesis maps each product term independently onto one multi-controlled gate. The arity of the widest gate is therefore fixed by the widest cube in the minimized expression, and no reduction in term count or literal count can lower it. This is the coupling that limits conventional synthesis: minimization operates on expression size, while cost is governed by expression width, and the two are not the same quantity.

Factorization breaks this coupling. Computing a shared factor once onto an auxiliary line and using that line as a control replaces one wide cube with several narrow ones. The circuit acquires more gates, but every gate is cheaper, and because cost increases superlinearly in arity the exchange is strongly favourable. The consequence is counterintuitive and central to this work: the factorized circuit is larger by every conventional structural measure and cheaper to execute.

## Factorized representation framework

The framework identifies two relationships among product terms and extracts the shared computation they expose.

**Containment.** When the literals of one term form a subset of another's, the smaller term is a common factor. For b'd and a'b'cd, the identity $X \oplus XY = X \cdot Y'$ gives $b'd \oplus a'b'cd = b'd(a'c)'$, so the shared factor is evaluated once and the residual a'c is computed onto an auxiliary line whose complement gates it.

**Complementary polarity.** When two terms share a factor but differ in the polarity of one or more literals, an XOR identity collapses the redundant control structure. The terms abcde' and a'be'f share be' and differ in the polarity of a, combining as $abcde' \oplus a'be'f = be'(acd \oplus a'f)$. A five-control and a four-control gate are replaced by gates of three, two and three controls. Expressions are realized through Product-Sum-EXOR gates[23-24], in which shared products are computed once and reused across related XOR contributions (Fig. 1).

**Auxiliary qubits.** Each extracted factor is computed onto an auxiliary line, which then acts as an additional control on the gate applying the factor; this control is charged in both quantum cost and T-count throughout. Lines are returned to $|0\rangle$ by measurement-based uncomputation, which requires no non-Clifford resources, and are reused across factor groups. Across all 64 functions evaluated, at most two auxiliary qubits are live at any point, independent of function size, from three-variable functions to hundred-variable functions with six hundred product terms.

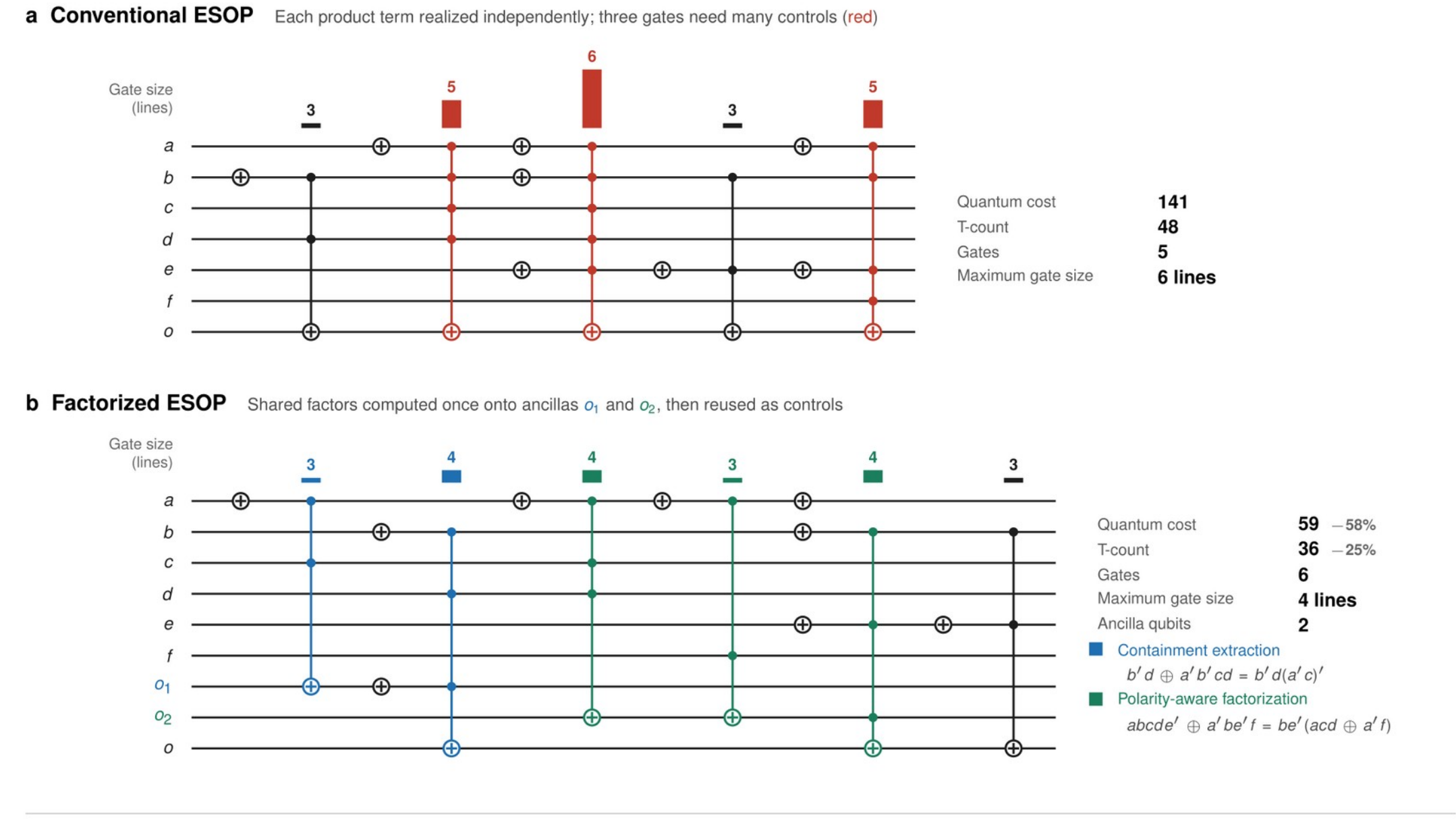

**Fig. 1 | Factorization reduces gate arity rather than gate count.** a, Conventional realization of the benchmark function con1f1, in which each of five product terms becomes an independent multi-controlled gate; three require many controls (red) and the widest spans six lines. b, The same function after containment-based factor extraction (blue) and polarity-aware factorization (green). Shared factors are computed onto auxiliary lines that then act as additional controls, so that no gate exceeds four lines. Badges give gate size in lines (controls plus target) and bar height is proportional to gate cost. The factorized circuit contains six gates rather than five, yet quantum cost falls from 141 to 59, T-count from 48 to 36, and the widest gate from five controls to three. Two auxiliary qubits are required.

## Results

We evaluated EXORCISM-5 on 64 single-output Boolean functions in three groups: 37 structured benchmarks from the RevLib reversible-circuit collection and the EPFL logic-synthesis suite[25-27]; 14 randomly generated functions of 100 input variables with 200 to 600 product terms; and 13 oracles constructed from quantum algorithms. Quantum cost was evaluated with the Maslov model and T-count with an AND-tree decomposition. Functional equivalence between each factorized representation and its originating expression was verified exhaustively for functions of 20 or fewer variables and by random sampling above that; every function passed.

Factorization reduced quantum cost for 56 of the 64 functions and left 8 unchanged, with no increase in quantum cost or T-count for any function at either stage (Fig. 2a, b). The median quantum-cost reduction was 42.7% and the median T-count reduction 28.6%.

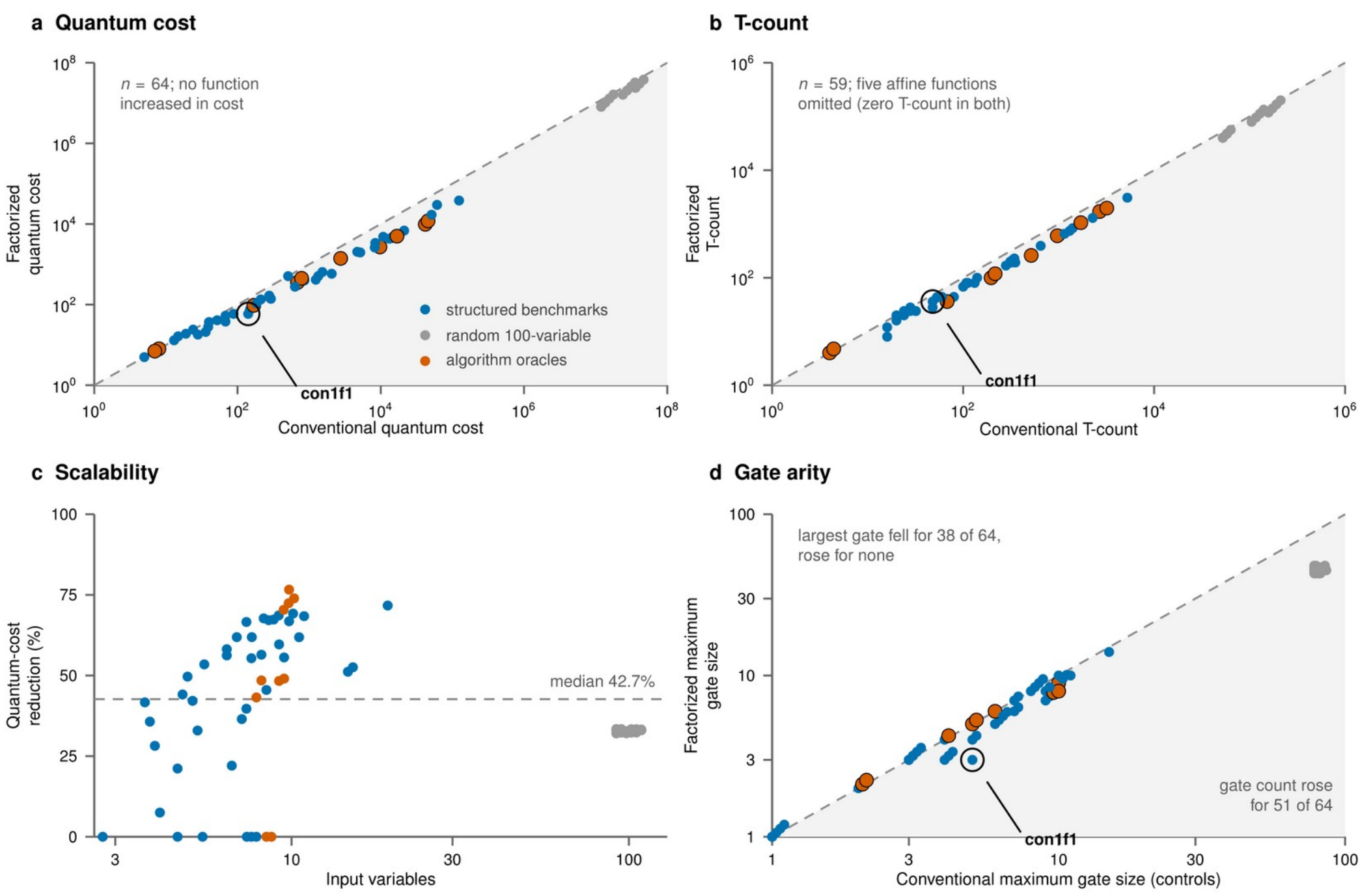

**Fig. 2 | Resource reductions across 64 Boolean functions.** Colour distinguishes three groups: structured benchmarks (blue), randomly generated 100-variable functions (grey), and oracles derived from quantum algorithms (vermillion). a, Factorized versus conventional quantum cost; the dashed line marks equality and the shaded region a reduction. No function increased in cost. The example worked on Fig. 1 is highlighted. b, Factorized versus conventional T-count for the 59 functions with non-zero conventional T-count; five affine functions require no non-Clifford resources in either representation and are omitted. c, Quantum-cost reduction against input-variable count, with median 42.7% (dashed line). The three groups separate clearly: oracles reduce most, random functions least. d, Factorized versus conventional maximum gate size. The largest gate fell for 38 of 64 functions and rose for none, while total gate count rose for 51 - the two move in opposite directions, the signature of exchanging few wide gates for many narrow ones.

**The reduction depends on structure, not size.** The three groups behave distinctly (Fig. 2c). Structured benchmarks gave a median reduction of 49.7%; algorithm oracles 61.9%; and randomly generated 100-variable functions only 32.9%. Random functions of this width contain essentially no containment relationships, because containment requires the polarities of some forty shared literals to coincide simultaneously, and consequently the containment stage finds no merges at all in that group. The eight functions showing no reduction are XOR-dominated or single-term and contain little multi-controlled logic to factor. That factorization leaves such functions exactly unchanged, rather than degrading them, indicates that the framework exploits existing structure without introducing overhead when none is present.

**Reduction arises from decreased gate arity.** Across the suite, total gate count increased for 51 of the 64 functions, while the size of the widest gate decreased by 38 and increased by none; the widest gate encountered fell from 86 controls to 48 (Fig. 2d). This is the signature of arity reduction rather than gate elimination, and it distinguishes the approach from conventional optimization, which targets gate or term count.

Monotonicity is guaranteed by construction. For T-count this is not only an empirical observation but follows from the two identities above: in each case, the T-count of the merged gates is no higher than that of the two independent gates they replace, for any shared factor and any residuals, and is strictly lower whenever the shared factor contains more than one literal (see Methods for the general form). No containment or polarity merge can therefore increase T-count for any Boolean function under this cost model. The guarantee is a property of the model rather than of an executed circuit: once auxiliary lines are uncomputed and multi-controlled gates decomposed into a Clifford+T basis, the accounting changes, and we quantify that difference below.

## Comparison with circuit optimization

The reductions above are measured on Boolean representations. To determine what survives translation into an executed circuit, and whether the same reductions are reachable by circuit-level rewriting, both realizations were exported to OpenQASM with multi-controlled gates decomposed into a Clifford+T basis and auxiliary lines uncomputed, then optimized with PyZX[28] using full graph reduction followed by circuit extraction. Of the 64 pairs, the 14 hundred-variable functions were not submitted, circuits of that width lying beyond the optimizer's practical range, and five large oracle instances were optimized separately under an extended limit and are reported below. Of the 45 pairs submitted under a 600-second limit, PyZX completed 39.

Two results follow. First, the advantage measured at the representation level is roughly halved once circuits are executed. Across the 34 completed benchmarks with non-zero T-count, factorization alone reduced T-count by a median of 16.7%, against the 33.5% obtained under the representation-level model. The difference is uncomputation, which an executed circuit must perform, and the representation-level model does not charge; in 3 of the 34 cases the factorized circuit carries a higher T-count than the conventional one before further optimization.

Second, factorization nonetheless improves what a state-of-the-art circuit-level optimizer achieves. PyZX applied to the conventional realization reduced T-count by a median of 42.3%, exceeding factorization alone. Applied to the factorized realization it reached a median of 53.4%, and produced the lower T-count for 24 of the 34 benchmarks against 5 for the conventional realization, with 5 ties; a two-sided sign test over the 29 decided cases gives p = 0.0005, the probability of a split at least this uneven arising by chance if the two realizations were equally likely to give the lower count. The median additional reduction attributable to factorization, measured after optimization, is 17.8%. Factorization therefore functions as a representation-level preprocessing step that exposes reductions the optimizer does not reach unaided, rather than as a competitor to circuit-level optimization.

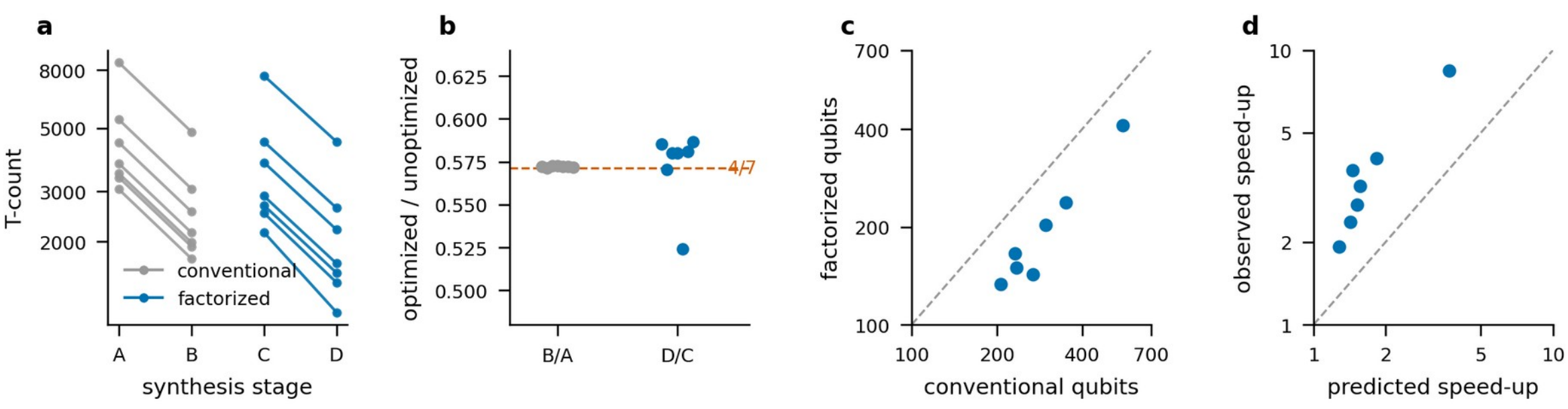

**Fig. 3 | Factorization improves circuit-level optimization on the largest completed comparisons.** Seven benchmark pairs whose PyZX optimization exceeded ten minutes. a, T-count at each of four stages: conventional (A), conventional optimized (B), factorized (C), factorized optimized (D) as a slopegraph; every trajectory from B to D decreases. b, Ratio of optimized to unoptimized T-count for both representations; values cluster tightly around a constant near 4/7, indicating the optimizer applies a fixed local rewrite independent of input structure. c, Qubit count, conventional against factorized; every point falls below the equality line, a reduction PyZX does not produce on its own. d, Observed PyZX speed-up on the factorized circuit against the speed-up predicted from circuit-size reduction alone; six of seven points lie above the diagonal. An eighth instance, the widest circuit attempted, is reported in the text but excluded from these panels: optimization of its conventional realization did not complete within the time limit, so no paired comparison is available.

On the largest completed comparisons, factorization improves circuit-level optimization consistently. Among the seven benchmarks whose PyZX optimization required more than ten minutes to complete, the largest instances attempted, up to 8,505 in baseline T-count — the factorized representation yielded a lower T-count than the conventional one after optimization in every case (Fig. 3a). PyZX alone reduced T-count by a median of 42.8%; applied to the factorized representation it reached a median of 51.0%, an additional reduction attributable to factorization of 247 – 1,021 T gates per instance. The ratio of optimized to unoptimized T-count was closely constant across both representations and across nearly an order of magnitude in circuit size (0.571 – 0.573 for the conventional realization, 0.524–0.587 for the factorized one; Fig. 3b), consistent with PyZX applying a fixed local rewrite rather than exploiting circuit-specific structure; the two reductions act on different aspects of the circuit and combine rather than compete. Qubit count fell in every instance, by a median of 32.2% (26.3 – 46.6%; Fig. 3c), a reduction the circuit-level optimizer does not produce. The optimizer also ran faster on the factorized circuits, by 1.9- to 8.4-fold (median 3.2-fold); this exceeded the reduction predicted from circuit size alone for six of the seven instances (Fig. 3d), indicating that factorization does not merely shrink the circuit PyZX must process but changes its structure in a way that is easier to optimize.

The two largest instances attempted are the output bits of $a^x$ mod 33. For the first, factorization reduced pre-optimization T-count by 10.0% and qubit count by 26.3%; after PyZX optimization, the factorized representation reached a T-count of 4,489 against 4,863 for the conventional representation optimized alone, a further 7.7% reduction, while the optimizer completed 1.9 times faster on the factorized circuit (35,745 s against 68,586 s). For the second, the widest circuit attempted at 663 conventional qubits, factorization reduced pre-optimization T-count from 10,129 to 8,764 (13.5%) and qubit count from 663 to 464 (30.0%). Optimization of the factorized circuit completed in 13.3 hours, reaching a T-count of 5,110; optimization of the conventional circuit did not complete within a 24-hour limit. This is the only instance among the eight largest attempted for which optimization of the conventional circuit failed to finish, indicating that the resource reduction from factorization can determine whether downstream optimization is tractable within a fixed budget, not merely how much it achieves. Because no optimized conventional T-count exists for this instance, it is excluded from the medians and paired comparisons reported above and from Fig. 3.

Factorization also reduces the width of the executed circuit. The factorized realization required fewer qubits than the conventional one for 39 of the 50 exported comparisons, unchanged for 10 and more for 1, with a median reduction of 28.1% among those that improved: for example 138 to 90 qubits for rd84f4_100, 88 to 56 for mux_185, and 67 to 37 for majority7. This reduction is not obtained by the circuit-level optimizer, which preserves qubit count. Since qubit footprint governs the physical size of a fault-tolerant implementation as directly as magic-state consumption governs its runtime, this is a distinct and complementary benefit.

## Application to algorithm oracles

Standard benchmarks establish behaviour on curated functions, but the Boolean functions a quantum algorithm actually synthesizes are oracles. We constructed oracles from three families: modular exponentiation bits $a^x$ mod N, which constitute the arithmetic core of Shor factoring; ripple-carry adder output bits, the primitive from which modular arithmetic is composed; and majority functions.

Across the 13 oracle instances the median quantum-cost reduction was 61.9% and the median T-count reduction 38.5%, exceeding the structured benchmark group. For the four non-degenerate modular-exponentiation bits the median reduction was 73.2% in quantum cost and 37.9% in T-count; two further bits of $a^x$ mod 15 are degenerate, their minimized expressions consisting of one and two cubes with no factorable structure, and are reported but excluded from the median. Adder output bits gave 48.4% and majority functions 68.6%.

Because each T gate consumes one distilled magic state, these reductions correspond to proportional reductions in magic-state consumption under the synthesis model assumed here. For the two measured output bits of $a^x$ mod 33, conventional synthesis requires 5,876 magic states and the factorized representation 3,680, a reduction of 37%. The circuits compute identical functions, verified exhaustively, and the saving arises entirely from the representation.

**Table 1 | Resource reductions on quantum algorithm oracles.**

| Oracle family | n | Quantum cost | T-count | Range (cost) |
|---|---|---|---|---|
| Shor modular exponentiation | 4 | 73.2% | 37.9% | 70.4-76.6% |
| Ripple-carry adder bits | 4 | 48.4% | 49.0% | 43.2-49.0% |
| Majority | 3 | 68.6% | 40.5% | 61.9-69.2% |
| All oracles | 13 | 61.9% | 38.5% | 0-76.6% |

Median reductions relative to conventional synthesis of the same function, by family. Two output bits of $a^x$ mod 15 are degenerate, their minimized expressions consisting of one and two product terms with no factorable structure; they are excluded from the Shor medians and retained in the deposited data.

These oracles are Boolean functions arising in quantum algorithms, synthesized here from their truth tables. They are not implementations of the algorithms themselves, which compose modular arithmetic hierarchically rather than synthesizing it from a truth table, and the instances evaluated are correspondingly small. The results establish that the exploitable structure is present in algorithmically derived functions and is not an artefact of curated benchmarks; they do not establish behaviour at cryptographically relevant scales.

## Scaling beyond truth-table synthesis

The results above are obtained by minimizing a truth table and factorizing the resulting expression, a route that is limited to functions of roughly a dozen variables because the table has 2n rows. The oracles a quantum algorithm executes are far wider than this, so a method confined to that regime cannot speak to the circuits that matter. We therefore asked whether the same two relationships can be extracted from a circuit that is composed rather than enumerated.

We generated modular exponentiation circuits directly from their arithmetic structure: a cascade of controlled modular multipliers, each a sequence of modular additions, each built from a ripple-carry adder. Generation is polynomial in modulus width. A 16-bit circuit, comprising 619,792 gates on 120 qubits with a T-count of 955,968, was produced in under one second, at a width where a truth table would require 65,536 rows for a single output bit and at 32 bits would require more than four billion (Fig. 4d). Measured T-count grows as the 2.79 power of the modulus width across the range tested. A composed circuit is a gate sequence rather than one flat expression, so factorization cannot be applied to it directly. We recovered expressions from it block by block, converting each block to its algebraic normal form by Mobius transform[17] and treating the resulting monomials as product terms. Because this conversion enumerates the inputs a block reads, block size governs whether it is possible at all: at a six-bit modulus a whole adder reads 39 wires and cannot be converted, whereas the majority and unmajority cells from which the adder is assembled read at most seven wires, independently of the modulus. The cell is therefore the unit at which the method applies, and this is a property of the construction rather than a choice.

The consequence is that a circuit of any width is built from a fixed inventory of cells. Across moduli of 4, 5, 6 and 8 bits the number of distinct cells was 133, 133, 140 and 135 respectively, while the number of cell instances grew from 12,663 to 79,841 (Fig. 4a). Factorizing each distinct cell once and weighting by its multiplicity therefore yields the resource reduction for the complete circuit, with no enumeration at circuit scale at any point. Applying the transformation to these cells reduced T-count by 6.5%, 7.0%, 7.0% and 6.7% respectively at the four widths, and quantum cost by 3.3%, 3.0%, 3.0% and 2.9% respectively (Fig. 4b). No cell increased in either measure, consistent with the monotonicity established above. Every one of the 541 distinct cells was verified equivalent to its originating block by exhaustive evaluation, and each extracted expression was checked against direct simulation of the block it came from; end-to-end verification is unavailable at these widths, so correctness is established cellwise.

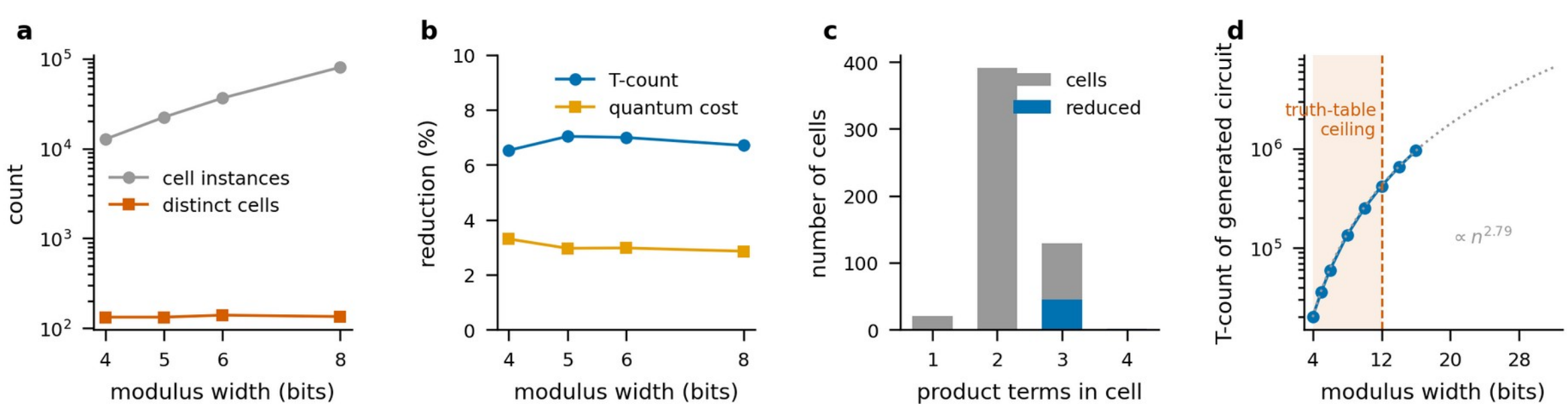

**Fig. 4 | Factorization applied to compositionally generated modular exponentiation.** Circuits for $a^x$ mod N were generated from their arithmetic structure at moduli of 4, 5, 6 and 8 bits and factorized cell by cell. a, Number of cell instances in the generated circuit against the number of distinct cells among them; instances grow with modulus width while the distinct inventory does not. b, T-count and quantum-cost reduction for the complete circuit, obtained by weighting each distinct cell by its multiplicity; both are flat in modulus width. c, Distinct cells grouped by the number of product terms they contain, with the subset whose T-count was reduced shown in colour; reduction is confined almost entirely to three-term cells, since factorization requires a pair of terms that share structure. d, Measured T-count of the generated circuit against modulus width, with the fitted power law and the width beyond which truth-table synthesis is not tractable; the compositional route passes that limit without enumeration.

The reduction is an order of magnitude smaller than on the benchmark suite, and the reason is visible in the cells themselves. Of the 541, twenty consist of a single product term and 391 of two; only 129 have three and one has four. Containment and complementary polarity act on pairs of terms that share structure, and a two-term cell rarely offers such a pair: 45 of the 46 cells that improved have three product terms, and none of the two-term cells improved at all (Fig. 4c). The transformation is not weaker at scale; it is applied to blocks that are individually too small to contain the structure it exploits. Factorizing across cell boundaries would expose that structure, at a conversion cost that grows exponentially in block width, and we did not attempt it here.

What the compositional route establishes is therefore a scaling property rather than a larger reduction. The per-cell inventory does not grow with the modulus, so under this observed invariance the measured per-cell reduction provides a width-independent estimate for the same compositional construction at any width: the 6.7% obtained at eight bits is the value the construction predicts at sixteen bits or at two thousand and forty-eight, a projection from an invariant cell inventory rather than a circuit we have factorized. The reductions of the preceding sections, by contrast, are measured on instances that cannot be constructed at those widths at all. The two results bound the method from opposite directions. Truth-table synthesis shows what the transformation achieves when an expression retains global structure; compositional synthesis shows what survives when it is confined to a modular block, and is a lower bound on what factorization of a composed circuit can deliver.

## Discussion

Boolean minimization does not define the ultimate representation for quantum realization. Minimization reduces expression size, whereas quantum cost is governed by expression width, and because these objectives differ a minimized expression can retain substantial exploitable structure. The reductions reported here quantify how much, obtained without altering the minimizer and without changing the function realized.

Factorized Boolean representation is therefore an optimization layer between logic minimization and reversible mapping, consuming the output of a conventional minimizer and producing input to a conventional mapper. Its relationship to circuit-level optimization is complementary rather than competitive: a ZX-calculus optimizer achieves a larger T-count reduction on its own than factorization does, but reaches a lower final count when applied to the factorized representation, and does not reduce qubit count at all. The effect is bounded by the present algebraic structure, which is substantial in arithmetic and symmetric functions and negligible in random ones, and the measured benefit is smaller in executed circuits than in the representations from which they are derived.

## Methods

### *Baseline expression generation*

Baseline exclusive-sum-of-products expressions were produced with EXORCISM-4 version 4.7 using its default minimization settings (-q 0, -v 0). The same expression serves as the baseline for cost comparison and as the input to factorization, so reported reductions isolate the effect of the representation change.

### *Cube representation*

Each product term is stored as a pair of 128-bit words: a value word in which a set bit denotes an input fixed to one, and a mask word in which a set bit denotes a don't-care position. Containment and overlap tests are therefore bitwise; literal counts are population counts, and the distance between two cubes over commonly fixed positions is a population count of their exclusive-or restricted to those positions.

### *Factorization procedure*

The containment mechanism (Stage 1) extends a cube-containment postprocessor introduced in earlier work[23-24], which identifies a contained cube and forms a PS implicant from it in the order cubes are encountered. The present implementation adds explicit cost evaluation of every candidate pair at each step and selects the greatest strictly positive reduction, as detailed below. The polarity-aware factorization (Stage 2) and its realization as a maximum-weight matching problem are introduced in this work.

Stage one extracts containment relationships. Cube A contains cube B when every literal B fixes is fixed identically in A and A fixes strictly more literals. The merge to apply at each step is chosen by explicit cost evaluation: for every candidate pair the quantum cost before and after merging is computed and the pair with the greatest reduction selected, with ties broken by smaller distance and then smaller residual literal count. Only merges with a strictly positive reduction are accepted. Each accepted merge reduces the term count by one, so the stage terminates after at most m-1 merges for m cubes.

Stage two extracts complementary-polarity relationships from terms that stage one left unfactored. For each valid pair the shared factor fixes positions where both terms fix the same literal and the residuals are the remainders; a pair is valid when the shared factor and both residuals are non-empty. Rather than merging greedily in sequence, this stage builds a weighted graph in which each term is a vertex and each valid pair an edge weighted by the number of shared literals, then accepts edges in decreasing weight provided neither endpoint is already matched. Each term therefore participates in at most one pair and the merges do not interfere.

***Algorithm 1 (containment).*** While candidate pairs remain: form every valid containment pair; for each, compute the cost of the two input terms and of their merged form, and take the difference; retain the pair with the greatest strictly positive reduction, breaking ties first by smaller cube distance and then by smaller residual literal count; replace that pair by its merged term. Terminate when no candidate yields a positive reduction. Because each accepted merge replaces two terms with one, at most m-1 merges occur.

***Algorithm 2 (complementary polarity).*** For every unordered pair of the terms stage one left unfactored, compute the shared factor as the literals fixed identically in both, and the residuals as the remainders; retain the pair as a weighted edge, of weight equal to the shared-factor literal count, when the factor and both residuals are non-empty. Sort edges by decreasing weight and accept each in turn provided neither endpoint is already matched, so that every term participates in at most one factorization. Apply the accepted pairs.

***Why T-count cannot increase.*** Write |L| for the literal count of a shared factor and |R| for that of a residual, using the notation introduced above. A containment merge replaces two independent gates of size |L| and |L|+|R| with two gates of size |R| and |L|+1; a polarity merge replaces gates of size |L|+|R1| and |L|+|R2| with gates of size |R1|, |R2| and |L|+1. Since T-count for an n-control gate is 4(n-1), the containment substitution reduces T-count by exactly 4(|L|-1) and the polarity substitution by exactly 4|L|, in both cases independent of the residuals and of which cubes are involved. The polarity reduction is strictly positive for any non-empty shared factor. The containment reduction is strictly positive for |L| > 1 and zero for |L| = 1, where the merge leaves T-count unchanged and is accepted only when it reduces quantum cost. Neither substitution can increase T-count for any shared factor and any residuals.

***Complexity.*** Cubes are stored as fixed-width bit vectors, so containment, distance and shared-factor tests are word-parallel and cost O([n/128]) machine operations rather than O(n). Cost evaluation is local: a candidate is scored from the two input terms and the merged term alone, not by recomputing the whole representation. Stage one therefore examines O($m^2$) candidates per iteration at O(1) amortized cost each, over at most m-1 iterations, giving O($m^3$) word operations. Stage two builds O($m^2$) edges, sorts them in O($m^2$ log m) and matches greedily in O($m^2$), giving O($m^2$ log m). Measured runtimes are consistent: on the 100-variable functions stage one required 4.6 ms at 200 terms, 29 ms at 400 and 112 ms at 600, and stage two remained below 5 ms throughout, so total synthesis time for the largest instance is well under a second.

### *Resource models*

Quantum cost was evaluated with the Maslov model, in which an n-control Toffoli is charged $2^{n+1} - 3$ elementary operations, corresponding to a decomposition using no auxiliary qubits. Negated literals are charged with two elementary operations each. T-count was evaluated as 4(n-1) per n-control gate, corresponding to a tree of measurement-assisted AND gates[29]. Both representations are evaluated with the identical cost function, including the charge for the auxiliary control, so reported reductions are properties of the representation rather than artefacts of the model.
Because the Maslov table assumes no auxiliary qubits are available for gate decomposition while the factorized realization uses them, both representations were additionally evaluated under a model granting auxiliary qubits to the baseline, in which an n-control gate decomposes into 2(n-2) Toffoli gates and costs O(n). Under this more conservative model the reduction persists: for con1f1, quantum cost falls from 92 to 50 (46%) compared with 141 to 59 (58%). Across the suite the median reduction is 27.7% under the auxiliary-qubit-aware model against 35.7% under the ancilla-free model, so the advantage does not arise from the exponential penalty the conventional model assigns to wide gates.

### *Auxiliary qubit accounting*

Auxiliary usage was measured rather than estimated. For each factored term the number of lines held simultaneously was recorded and the peak across the circuit reported. Lines are returned to $|0\rangle$ by measurement-based uncomputation between factor groups, permitting reuse; the peak simultaneous requirement across all 64 functions is two.

### *Equivalence verification*

Every factorized representation was checked against its originating expression by direct evaluation: exhaustively over all $2^n$ assignments for functions of 20 or fewer variables, and over 20,000 uniformly random assignments above that. All 64 functions passed. For the 100-variable functions random sampling covers a negligible fraction of the input space and constitutes strong evidence rather than proof; a failure is conclusive, a pass is not.

### *Comparison with circuit optimization*

Both realizations were exported to OpenQASM 2.0 with multi-controlled gates decomposed into a Clifford+T basis and auxiliary lines uncomputed, so that exported circuits are executable, and their T-counts exceed those of the representation-level model. Each circuit was optimized with PyZX 0.10.5 by converting to a ZX graph, applying full graph reduction, and extracting a circuit, with T-count measured on the extracted basic-gate circuit. The conventional and factorized realizations received identical treatment. Optimization ran in a separate process under a 600-second wall-clock limit. The 14 hundred-variable functions were not submitted, circuits of that width lying beyond the optimizer's practical range. Of the 45 pairs submitted, 6 exceeded the limit and are excluded: sym10_d_100, 9sym_d_100, life_d_100, max46_d_100, ryy6_198 and majority11, each timing out on both realizations. Five further oracle instances were optimized under an extended limit of 8 to 24 hours and are reported separately in Fig. 3. All attempts, completed and timed out, are listed in the deposited comparison files. Median PyZX runtime on the completed pairs was 1.4 s.

### *Benchmark suite and oracle construction*

The suite comprises 37 structured benchmarks drawn from RevLib and the EPFL logic-synthesis suite[25-26], 14 randomly generated 100-variable functions with 200, 400 or 600 product terms, and 13 oracles. Oracles were specified as Boolean functions and converted to exclusive-sum-of-products form through the same EXORCISM-4 path as the benchmarks. Modular exponentiation oracles are the individual output bits of $a^x$ mod N for (a, N) in {(7,15), (2,21), (5,33)}; adder oracles are the high-order sum and carry bits of 4- and 5-bit ripple-carry addition; majority oracles are the n-input majority functions for n in {7, 9, 11}. The 14 synthetic functions were generated by drawing each of the 100 positions of each product term independently and uniformly from the three symbols (don't-care, complemented, uncomplemented), so that a variable appears in a term with probability two-thirds and with either polarity equally often; duplicate terms were rejected, since a repeated term cancels under Exclusive-OR. Instances use the term counts 200, 400 and 600 with seed equal to the instance index. The distributed instances match this model over 560,000 positions, with symbol frequencies of 33.40%, 33.23% and 33.37% and a chi-square of 2.9 against a uniform null (critical value 9.21 at $p = 0.01$, 2 d.f.).

### *Compositional oracle generation*

Modular exponentiation circuits were generated as a cascade of controlled modular multipliers over a classical modulus, each multiplier a sequence of controlled modular additions[30] and each addition a ripple-carry adder in the Cuccaro formulation[31], using majority and unmajority cells with the two-controlled-not unmajority variant. Registers are one bit wider than the modulus so that the leading position records the sign after a subtraction; a modulus that does not satisfy this condition is rejected rather than silently mis-synthesized. In-place multiplication uses the standard multiply, controlled-swap, inverse-multiply sequence with the modular inverse computed classically, which requires the multiplier to be coprime to the modulus.

Each generated circuit was verified before any factorization was applied. The adder was checked exhaustively for widths up to eight bits over both carry-in values, and at 12, 16, 20, 24, 32, 48 and 64 bits against edge cases and 40,200 random operand pairs, confirming the sum, the carry-out, preservation of the addend and restoration of the carry ancilla. Verification was itself validated by injecting seven classes of fault into the circuit, including a dropped gate, a lost control, a reversed cell order and an off-by-one carry wire; all seven were detected while the unmodified circuit passed. The modular adder was verified exhaustively over every modulus and every operand pair below it, and the in-place multiplier over every modulus, every coprime multiplier and both control values, in each case confirming that all ancillas return to zero.

Expressions were recovered from each block by evaluating it over all assignments to the wires it reads and applying a Mobius transform[17] to obtain the positive-polarity Reed-Muller form, whose monomials are the product terms. Blocks were taken to be the three-gate majority and unmajority cells, which read at most seven wires at any modulus width; coarser segmentations were measured and rejected, a whole adder reading 39 wires at a six-bit modulus. Identical cells were merged and their multiplicity recorded, so that each distinct cell is factorized once and the circuit total obtained by weighting. Every extracted expression was checked against direct simulation of its block over all assignments, and factorized cells were checked by the same equivalence procedure used for the benchmark suite. Resource counts use the models defined above, so the figures are comparable with those reported for the benchmarks.

## Data availability

Benchmark and oracle specifications, generated expressions at every stage, the per-function resource table, and the circuit-level comparison results including all timed-out attempts are available at https://github.com/mehulshah225/Research_Project. The structured benchmarks derive from the RevLib reversible-circuit collection (https://revlib.org) and the EPFL logic-synthesis benchmark suite (https://github.com/lsils/benchmarks).

## Code availability

The EXORCISM-5 implementation, resource-analysis tool and equivalence checker are available at https://github.com/mehulshah225/Research_Project.

## Funding

This research received no specific grant from any funding agency in the public, commercial or not-for-profit sectors.

## Author contributions

M.P. and R.F. introduced Product-Sum-EXOR gates and a cube-containment postprocessor for reversible synthesis in earlier work[23-24], which this work builds on and extends. M.S. introduced cost-based candidate selection for the containment stage, devised and implemented the polarity-aware factorization stage and its realization as a maximum-weight matching problem, built the resource-analysis and equivalence-verification tooling, generated the benchmark and oracle suites, performed all evaluations including the comparison with circuit-level optimization, and prepared the figures. M.P. supervised the work. All authors discussed the results and contributed to the manuscript.

## Competing interests

The authors declare no competing interests.

## Additional information

Correspondence and requests for materials should be addressed to M.S. (mehul@pdx.edu).